\input{aipcheck}

\documentclass[final]{aipproc}

\layoutstyle{6x9}

\newcommand{\ee}{\end{equation}} 
\newcommand{\be}{\begin{equation}} 
\newcommand{\ec}{\end{center}} 
\newcommand{\bc}{\begin{center}} 
\newcommand{\eea}{\end{eqnarray}} 
\newcommand{\bea}{\begin{eqnarray}} 
\newcommand{\bd}{\begin{description}} 
\newcommand{\ed}{\end{description}} 
\newcommand{\bi}{\begin{itemize}} 
\newcommand{\ei}{\end{itemize}}

\def\spose#1{\hbox to 0pt{#1\hss}}
\def\ltapprox{\mathrel{\spose{\lower 3pt\hbox{$\mathchar"218$}}
 \raise 2.0pt\hbox{$\mathchar"13C$}}}
\def\gtapprox{\mathrel{\spose{\lower 3pt\hbox{$\mathchar"218$}}
 \raise 2.0pt\hbox{$\mathchar"13E$}}}

\begin{document}

\title{Feynman gauge on the lattice: \\ new results and perspectives}

\classification{11.15.Ha 12.38.-t 12.38.Gc 14.70.Dj}
\keywords{Feynman gauge, Lattice gauge theory, Green's functions}

\author{Attilio Cucchieri}{
  address={Instituto de F\'\i sica de S\~ao Carlos, Universidade de S\~ao Paulo, \\[1mm]
           Caixa Postal 369, 13560-970 S\~ao Carlos, SP, Brazil \\[2mm]}
}

\author{Tereza Mendes}{
  address={Instituto de F\'\i sica de S\~ao Carlos, Universidade de S\~ao Paulo, \\[1mm]
           Caixa Postal 369, 13560-970 S\~ao Carlos, SP, Brazil \\[2mm]}
}

\author{Gilberto M.\ Nakamura}{
  address={Instituto de F\'\i sica de S\~ao Carlos, Universidade de S\~ao Paulo, \\[1mm]
           Caixa Postal 369, 13560-970 S\~ao Carlos, SP, Brazil \\[2mm]}
}

\author{Elton M.\ S.\ Santos}{
  address={Instituto de F\'\i sica de S\~ao Carlos, Universidade de S\~ao Paulo, \\[1mm]
           Caixa Postal 369, 13560-970 S\~ao Carlos, SP, Brazil \\[2mm]}
  ,altaddress={Instituto de Educa\c c\~ao, Agricultura e Ambiente, Campus Vale do Rio Madeira, \\[1mm]
               Universidade Federal do Amazonas, 69800-000 Humait\'a, AM, Brazil}
}

\begin{abstract}
We have recently introduced a new implementation of the Feynman gauge on
the lattice, based on a minimizing functional that extends in a natural
way the Landau-gauge case, while preserving all the properties of the
continuum formulation. The only remaining difficulty with our approach
is that, using the standard (compact) discretization, the gluon field is
bounded, while its four-divergence satisfies a Gaussian
distribution, i.e.\ it is unbounded. This can give rise to convergence
problems when a numerical implementation is attempted. In order to overcome
this problem, one can use different discretizations for the gluon field, or
consider an SU($N_c$) group with sufficiently large $N_c$. Here we discuss
these two possible solutions.
\end{abstract}

\maketitle

\section{Introduction}

The behavior of Green's functions in the infrared limit of Yang-Mills theories should give us
some insights into the low-energy properties of these theories. Since these functions depend on
the gauge condition, considering different gauges could help us gain a better understanding
of the (non-perturbative) low-energy hallmarks of QCD, such as color confinement. In the last 20
years, several groups have used lattice simulations to study propagators and vertices of Yang-Mills
theories in Landau gauge \cite{Cucchieri:2010xr}, Coulomb gauge \cite{Cucchieri:2006hi,Burgio:2008jr,
Greensite:2009eb}, $\lambda$-gauge (a gauge that interpolates between Landau and Coulomb)
\cite{Cucchieri:1998ta,Cucchieri:2007uj} and maximally Abelian gauge \cite{Bornyakov:2003ee, 
Mendes:2006kc}.

On the other hand, until recently, the numerical gauge fixing for the linear covariant gauge ---
which is a generalization of Landau gauge --- was not satisfactory \cite{Giusti:1996kf,Giusti:1999wz,
Giusti:1999im,Giusti:1999cw,Giusti:2000yc,Giusti:2001kr,Cucchieri:2008zx,Mendes:2008ux}. In Ref.\
\cite{Cucchieri:2009kk} we have introduced a new implementation of the linear covariant gauge on
the lattice that solves most problems encountered in earlier implementations (see \cite{Cucchieri:2009kk,
Cucchieri:2010ku} for a short review of early works). As explained in the abstract, the only
problem still affecting our method, as well as any formulation of the linear covariant
gauge on the lattice, is due to the fact that the gluon field $A^a_\mu(x)$ is {\em bounded} in the usual
compact formulation of lattice Yang-Mills theories. On the contrary, the functions $\Lambda^b(x)$
satisfy a Gaussian distribution, i.e.\ they are {\em unbounded}. Thus, one has to deal with convergence
problems \cite{Rank} when numerically fixing the gauge condition
\begin{equation}
\partial_{\mu} A^b_\mu(x)=\Lambda^b(x) \; .
\label{eq:lincov}
\end{equation}
Since the real-valued functions $\Lambda^b(x)$ are generated using a Gaussian distribution with
width $\sqrt{\xi}$, it is clear that this problem becomes more severe when $\xi$ is larger and/or
when the lattice volume is larger. Here we discuss two possible solutions for this problem,
namely we consider different discretizations for the gluon field or a gauge group SU($N_c$) with
sufficiently large $N_c$.

\section{Linear covariant gauge on the lattice}

We want to impose the gauge condition (\ref{eq:lincov}) on the lattice. Landau gauge, which
corresponds to the case $\Lambda^b(x)=0$, is obtained on the lattice by minimizing the functional
\begin{equation}
{\cal E}_{LG}[U^{g}] = - \mbox{Tr} \sum_{\mu, x} g(x) U_{\mu}(x) g^{\dagger}(x+e_{\mu}) \; .
\label{eq:ELG}
\end{equation}
Here $U_{\mu}(x)$ are link variables and $g(x)$ are site variables, both belonging to the SU($N_c$)
group. The sum is taken over all lattice sites $x$ and directions $\mu$. Also, $\mbox{Tr}$ indicates
trace in color space. For the linear covariant gauge we can look for a minimizing functional of the
type ${\cal E}_{LCG}[U^{g}, g, \Lambda]$. If one recalls that solving the system of equations
$B \phi = c$, where $B$ is a matrix and $\phi$ and $c$ are vectors, is equivalent to
minimizing the quadratic form $ \frac{1}{2} \phi B \phi - \phi c $, then it is obvious
that in our case we should look for a minimizing functional of the type \cite{Cucchieri:2009kk}
\begin{equation}
{\cal E}_{LCG}[U^{g}, g, \Lambda] \sim {\cal E}_{LG}[U^{g}] - g \Lambda \; .
\end{equation}
Indeed, the lattice linear covariant gauge condition can be obtained by minimizing\footnote{One
should stress that, in the minimization process, the link variables $U_{\mu}(x)$ get gauge-transformed
to $g(x) U_{\mu}(x) g^{\dagger}(x+e_{\mu})$, while the $\Lambda^b(x)$ functions do not get
modified.}
\begin{equation}
{\cal E}_{LCG}[U^{g}, g, \Lambda] \; = \; {\cal E}_{LG}[U^{g}]
                   \, + \, \Re \; Tr \sum_x \,  i\, g(x) \, \Lambda(x) \; ,
\end{equation}
where ${\cal E}_{LG}[U^{g}]$ is defined above in Eq.\ (\ref{eq:ELG}) and $ \Re $ indicates real part.
This can be checked by considering a one-parameter subgroup $ g(x,\tau) \, = \, \exp \left[ i \tau
\gamma^{b}(x) \lambda^{b} \right] $. Here we indicate with $\lambda^{b}$ a basis for the SU($N_c$)
Lie algebra and with $\gamma^{b}(x)$ any real-valued functions. Indeed, the stationarity condition implies
the lattice linear covariant gauge condition\footnote{Note that periodic boundary conditions yields
$ \sum_x \Lambda^{b}(x) = 0$. This equality has to be enforced explicitly, within machine precision,
when the functions $ \Lambda^{b}(x)$ are generated.}
\begin{equation}
\sum_{\mu} \, A^b_\mu(x) \,-\, A^b_\mu(x-e_{\mu}) \, = \, \Lambda^b(x) \; .
\label{eq:lincovlatt}
\end{equation}
Also, the second variation (with respect to the parameter $\tau$) of the term $\,i\, g(x) \, \Lambda(x)\,$
is purely imaginary and it does not contribute to the Faddeev-Popov matrix ${\cal M}$, i.e.\ ${\cal M}$
is a discretized version of the usual Faddeev-Popov operator $- \partial \cdot D$. Let us note that
having a minimizing functional for the linear covariant gauge implies that the Faddeev-Popov operator
${\cal M}$ is positive-definite and that the set of its local minima defines the first Gribov region
$\Omega$.\footnote{This region has been studied analytically in \cite{Sobreiro:2005vn}, for a small value of $\xi$, but a
similar numerical study is still lacking.}

It is interesting to note that one can interpret the Landau-gauge functional $ {\cal E}_{LG}[U^{g}] $
as a spin-glass Hamiltonian \cite{Marinari:1991zv} for the spin variables $g(x)$ with a random interaction given by
$U_{\mu}(x)$. Then, our new functional corresponds to the same spin-glass Hamiltonian when a random
external magnetic field $ \Lambda(x) $ is applied.

Note also that the functional ${\cal E}_{LCG}[U^{g}, g, \Lambda]$ is linear in the gauge transformation
$\{ g(x) \}$. Thus, one can easily extend to the linear covariant gauge the gauge-fixing algorithms
usually employed in the Landau case \cite{Cucchieri:1995pn,Cucchieri:1996jm,Cucchieri:2003fb}. We refer
the reader to References \cite{Cucchieri:2009kk,Cucchieri:2010ku} for tests of convergence of the numerical
gauge fixing. There we have also checked that the quantity $D_l(p^2) p^2$, where $D_l(p^2)$ is the
longitudinal gluon propagator, is approximately constant for all cases
considered, as predicted by Slavnov-Taylor identities. This verification failed in previous formulations
of the lattice linear covariant gauge \cite{Giusti:2000yc,Mendes:2008ux}.

\begin{table}
\begin{tabular}{cccc}
\hline
 $\xi$ & stand.\ & angle & stereog.\ \\
 \hline
 $0.01$ & 2.2 & 2.2 & 2.2 \\
 $0.05$ & 2.2 & 2.2 & 2.2 \\
 $0.1$  & 2.2 & 2.2 & 2.2 \\
 $0.5$  & 2.8 & 2.6 & 2.5 \\
 $1.0$  & --- & 3.0 & 2.5 \\
 \hline
\end{tabular}
\caption{Smallest value of $\beta$ for which the numerical gauge-fixing
algorithm showed convergence. Results are reported for the three different
discretizations and for five different values of the gauge parameter
$\xi$.}
\label{tab:a}
\end{table}

\section{Discretization effects}

As explained above, the standard discretization of the gluon field $A^a_\mu(x)$ is bounded.
Since the functions $\Lambda^b(x)$ are generated using a Gaussian distribution, it is clear that
Eq.\ (\ref{eq:lincovlatt}) cannot be satisfied if $\Lambda^b(x)$ is too large. A
possible solution to this problem is to use different discretizations of the gluon field.
We did some tests in the SU(2) case using the angle (or logarithmic) projection \cite{Amemiya:1998jz}
and the stereographic projection \cite{vonSmekal:2007ns} (for a slightly different implementation
of the stereographic projection see also \cite{Gutbrod:2004qp}). Note that, in the latter case, the gluon field
is unbounded even for a finite lattice spacing $a$.

In particular, considering the standard discretization,
the angle projection and the stereographic projection for the lattice volume $V=8^4$, gauge parameter
$\xi=0.01, 0.05,0.1,0.5,1.0$ and lattice coupling $\beta=2.2,2.3,\ldots,2.9,3.0$ we checked (using for
the numerical gauge fixing the so-called Cornell method \cite{Cucchieri:1995pn,Cucchieri:1996jm,Cucchieri:2003fb})
in which cases we were able to fix the covariant gauge condition effectively. Our results, reported in Table
\ref{tab:a}, clearly show that the angle projection is already an improvement compared to the standard
discretization and that the best convergence is obtained when using the stereographic projection.

\section{Continuum Limit}

Note \cite{Mendes:2008ux} that the continuum relation
\begin{equation}
\partial_{\mu} A^b_\mu(x)=\Lambda^b(x)
\end{equation}
can be made dimensionless --- working in a generic $d$-dimensional space --- by
multiplying both sides by $a^2 g_0$. Since $\beta = 2 N_c / (a^{4-d} g_0^2)$
[in the SU($N_c$) case], we have that the lattice quantity
\begin{equation}
\frac{\beta / (2 N_c)}{2\xi} \sum_{x, b} \left[ a^2 g_0 \Lambda^b(x)\right]^2
\, = \, \frac{1}{2\sigma^2} \sum_{x, b} \left[ a^2 g_0 \Lambda^b(x)\right]^2
\end{equation}
becomes
\begin{equation}
 \frac{1}{2\xi} \frac{1}{a^{4-d} g_0^2} \int \frac{d^dx}{a^d}
     \sum_b \left[a^2 g_0 \Lambda^b(x)\right]^2 \;  = \;
     \frac{1}{2\xi} \int d^dx \sum_b \left[\Lambda^b(x)\right]^2 
\end{equation}
in the formal continuum limit. Thus, if we consider a gauge parameter $\xi$ in the
continuum, the lattice quantity $a^2 g_0 \Lambda^b(x)$ is generated from a Gaussian
distribution with width $\sigma=\sqrt{2 N_c \xi/\beta}$, instead of a width
$\sqrt{\xi}$.

\begin{table}
\begin{tabular}{ccccc}
\hline
 $N_c$ & $\;\beta_1\;$ & $\;\beta_2\;$ & $\;\beta_3\;$ & $\;\beta_4\;$ \\
 \hline
 2     & 3.0       & 2.485     & 2.295     & 2.44      \\
 3     & 6.75      & 6.67      & 6.07      & 5.99      \\
 4     & 12.0      & 12.59     & 11.43     & 10.97     \\
 \hline
\end{tabular}
\caption{Values of the lattice coupling $\beta$ considered
for the gauge groups SU(2), SU(3) and SU(4) with a gauge parameter
$\xi = 1$.}
\label{tab:b}
\end{table}

Note that $\sigma=\sqrt{\xi}$ if $\beta = 2 N_c$ and that for $\beta < 2 N_c$ the
lattice width $\sigma$ is larger than the continuum width $\sqrt{\xi}$, making the convergence
problem discussed above more severe. Thus, in the
SU(2) case, one has $\sigma = \sqrt{\xi}$ only for $\beta = 4$, corresponding to a
lattice spacing $a \approx 0.001$ fm. On the contrary, in the SU(3) case, one has
$\sigma = \sqrt{\xi}$ for $\beta = 6$, corresponding to $a = 0.102$ fm. Also, for a
fixed t'Hooft coupling $g_0^2 N_c$, we have $\beta \propto N_c^2$
and $\sigma \propto \sqrt{1/N_c}$. This suggests that simulations for the linear covariant gauge
are probably easier in the SU($N_c$) case for large $N_c$.

In order to test this hypothesis we simulated the SU(2), SU(3) and SU(4) cases for a
gauge parameter $\xi=1$ and lattice volumes $V=8^4, 16^4, 24^4, 32^4$ for the values of
$\beta$ reported in Table \ref{tab:b}. They correspond, respectively, to a t'Hooft coupling
$g_0^2 N_c = 8/3$ ($\beta_1$), to a plaquette average value of about 0.65 ($\beta_2$) and
of about 0.6 ($\beta_3$) and to a string tension (in lattice units) of about $a^2 \sigma =
0.044$ ($\beta_4$), giving $a \approx 0.09$ fm. In Table \ref{tab:c} we present, for each
pair (SU($N_c$),$\;V$), the values of the lattice coupling $\beta$ for which the gauge-fixing
algorithm showed a numerical convergence.\footnote{For these tests we used the standard
overrelaxation algorithm \cite{Cucchieri:1995pn,Cucchieri:1996jm,Cucchieri:2003fb}.} One
clearly sees that the situation improves when the number of colors $N_c$ is larger.

\section{Conclusions}

We have recently introduced a minimizing functional for the linear covariant gauge which is a
simple generalization of the Landau-gauge functional. The new approach solves most problems
encountered in earlier implementations and ensures a good quality for the gauge fixing.
Here we have shown that, by using different discretizations for the gluon field (such as
the angle projection) and by considering a gauge group SU($N_c$) with $N_c$ sufficiently large
--- i.e.\ $N_c = 4$ or maybe even $N_c = 3$ --- one should be able to do simulations for
$\xi = 1$ (Feynman gauge) and for large lattice volumes (in physical units). Let us note that
a numerical study of the infrared behavior of propagators and vertices at $\xi \neq 0$ could
provide important inputs for analytic studies based on Dyson-Schwinger equations \cite{Alkofer:2003jr,
Aguilar:2007nf}. Moreover, it has been proven \cite{Binosi:2002ft,Binosi:2003rr,Binosi:2009qm}
that the background-field Feynman gauge is equivalent (to all orders in perturbation theory)
to the pinch technique \cite{Binosi:2009qm,Cornwall:1981zr}. Thus, numerical studies using the
Feynman gauge, which corresponds to the value $\xi = 1$, could allow a nonperturbative evaluation
of the gauge-invariant off-shell Green functions of the pinch technique \cite{Cornwall:2009as}.

\begin{table}
\begin{tabular}{ccccc}
\hline
                  & $8^4$ & $16^4$ & $24^4$ & $32^4$ \\
 \hline
 SU(2)  & $\beta_1$, $\beta_2$ &  ---   &  ---   &  ---   \\
 SU(3)  & all      & $\beta_1$, $\beta_2$ & $\beta_1$, $\beta_2$ & $\beta_1$, $\beta_2 \; ^{*}$   \\
 SU(4)  & all  & all  & all  & $\beta_1$, $\beta_2$, $\beta_3$ \\
 \hline
\end{tabular}
\caption{Values of $\beta$ for which the numerical gauge-fixing algorithm
showed convergence. Results are reported
for three different gauge groups and four different
lattice volumes. In all cases the gauge parameter $\xi$ was
1 (Feynman gauge). $^{*}$[In these two cases only a few configurations have been considered and
more tests are needed.]}
\label{tab:c}
\end{table}

\begin{theacknowledgments}
This work has been partially supported by the Brazilian agencies FAPESP, CNPq and
CAPES. In particular, support from FAPESP (under grant \# 2009/50180-0) is acknowledged.
\end{theacknowledgments}

\bibliographystyle{aipproc}   

\end{document}